\documentclass[preprint]{sigchi}

\usepackage{balance}  
\usepackage{graphics} 
\usepackage{txfonts}
\usepackage{times}    
\usepackage[pdftex]{hyperref}
 \usepackage{url}      
\usepackage{color}
\usepackage{textcomp}
\usepackage{booktabs}
\usepackage{ccicons}
\usepackage{todonotes}
\usepackage[shortlabels]{enumitem}
\usepackage{array}
\newcolumntype{L}[1]{>{\raggedright\let\newline\\\arraybackslash\hspace{0pt}}m{#1}}
\newcolumntype{C}[1]{>{\centering\let\newline\\\arraybackslash\hspace{0pt}}m{#1}}
\newcolumntype{R}[1]{>{\raggedleft\let\newline\\\arraybackslash\hspace{0pt}}m{#1}}

\makeatletter
\def\url@leostyle{%
  \@ifundefined{selectfont}{\def\UrlFont{\sf}}{\def\UrlFont{\small\bf\ttfamily}}}
\makeatother
\def\pprw{8.5in}
\def\pprh{11in}

\definecolor{linkColor}{RGB}{6,125,233}
\hypersetup{%
  pdftitle={SIGCHI Conference Proceedings Format},
  pdfauthor={LaTeX},
  pdfkeywords={SIGCHI, proceedings, archival format},
  bookmarksnumbered,
  pdfstartview={FitH},
  colorlinks,
  citecolor=black,
  filecolor=black,
  linkcolor=black,
  urlcolor=linkColor,
  breaklinks=true,
}

\begin{document}

\title{SentenceRacer: A Game with a Purpose for \\ Image Sentence Annotation}

\numberofauthors{1}
\author{%
\alignauthor{Kenji Hata\thanks{These authors contributed equally to the publication.}, Sherman Leung\footnotemark[1], Ranjay Krishna, Michael S. Bernstein, Li Fei-Fei\\
     \affaddr{Stanford University}\\
     \email{\{kenjihata, sherman, rak248, msb, feifeili\}@cs.stanford.edu}
     }
}

\maketitle
\begin{abstract}
Recently datasets that contain sentence descriptions of images have enabled models that can automatically generate image captions. However, collecting these datasets are still very expensive. Here, we present SentenceRacer, an online game that gathers and verifies descriptions of images at no cost. Similar to the game hangman, players compete to uncover words in a sentence that ultimately describes an image. SentenceRacer both generates and verifies that the sentences are accurate descriptions. We show that SentenceRacer generates annotations of higher quality than those generated on Amazon Mechanical Turk (AMT). 
\end{abstract}


\section{Introduction}
The ability of describing images with sentences has numerous applications like helping the visually impaired independently browse the Internet. Recently, with competitions like Microsoft COCO's Image Captioning \cite{coco}, there has been an increased interest in the task of automatic image description generation \cite{google}. With this interest, there is a dire need for large scale datasets that can be used for training these sentence generation models. Datasets like COCO \cite{coco} and Flickr30M \cite{flickr} have been collected by crowdsourcing the description task to human workers on Amazon Mechanical Turk. Once the sentences are generated by one crowd worker, both \cite{coco, flickr} send their sentences to additional crowd workers to verify the accuracy of the sentences. The biggest bottleneck in growing these datasets to a much larger scale has been the cost of generating these sentences and verifying their accuracy. 

Humans are strikingly proficient at "filling in the blanks" --- whether it be crosswords, hangman, or Wheel of Fortune. We enjoy partially filled puzzles because of the feeling of simultaneously knowing and not knowing the full answer \cite{crossword}. Previous research by von Ahn and Dabbish show gamification to be an effective vehicle for labeling images in datasets for free \cite{gwap1}. However, such games were only limited to single word annotations \cite{gwap1}. This paper explores another game mechanism in order to generate more complex, full-sentence annotations. We propose a gamification method to crowdsource sentence annotations of images by having players write sentences for other players to guess. Additionally, we show that there exists a direct correlation between the accuracy of the sentence description and a player's ability to guess the sentence.

Motivated to reduce the cost of collecting a large image captioning dataset, we present a game that achieves the following: \begin{enumerate}\item Generates sentence descriptions of images \item Verifies that these sentences are accurate \item Captures sentences of better quality than those collected by Amazon Mechanical Turk (AMT). \end{enumerate}
\begin{figure}[t]
\centering
\includegraphics[width=3.3in]{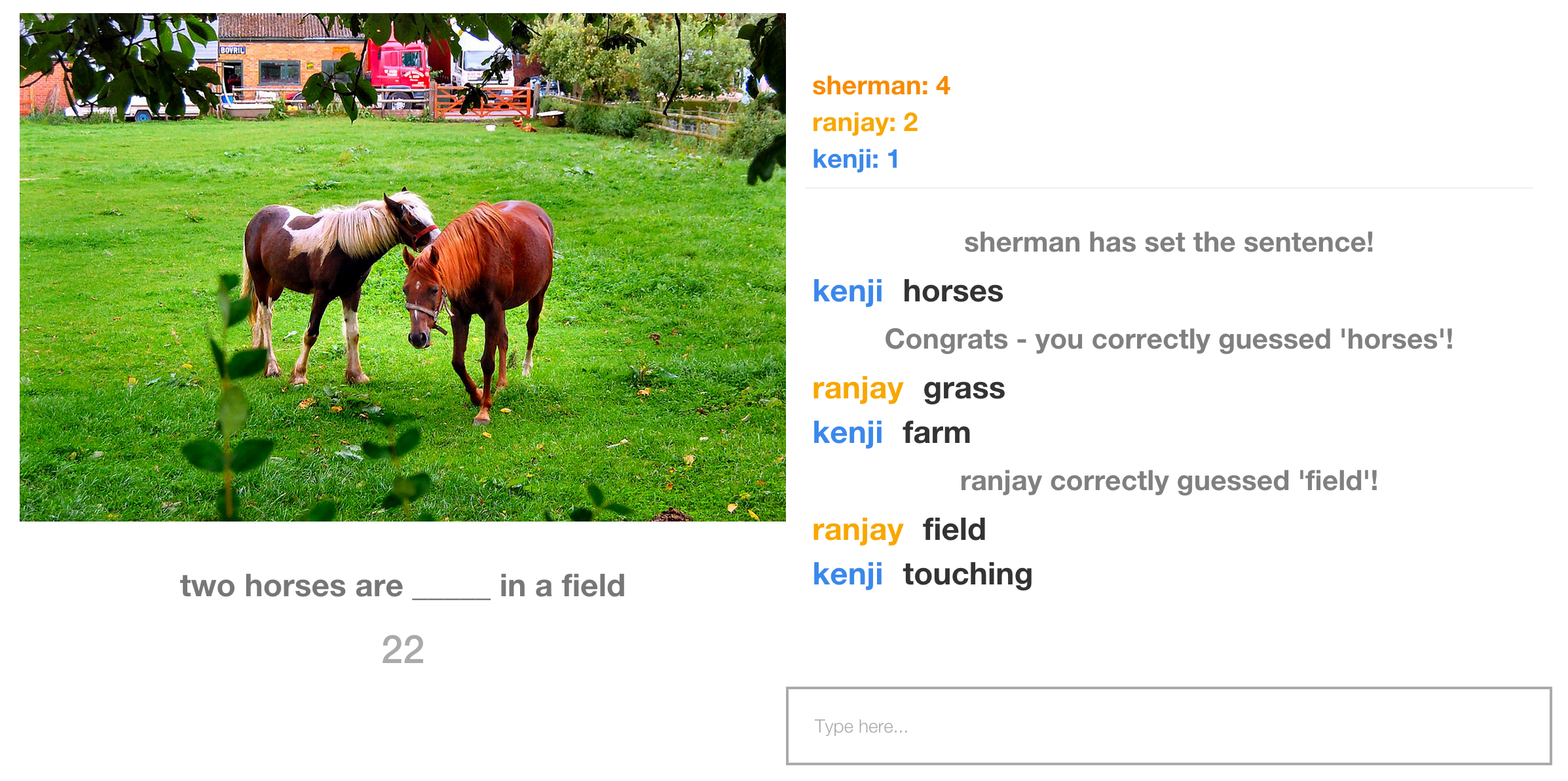}
\caption{A screenshot of SentenceRacer's interface. The left side displays the image and the state of the verified words so far. The right side displays the chat interface and the scoreboard for guessing.} 
\label{fig-example2}
\end{figure}
\section{System Description}
SentenceRacer is played with a minimum of three players. Each round of the game rotates a leader position. The leader sets a sentence describing the image for other players to guess. After eliminating stop words from the sentence, we allow all players to see guesses made by other players while blocking the leader's communication with the guessers. Players only have limited time to guess the words set by the leader. A correct guess rewards both the guesser and the leader with points and reveals the guess' position in the sentence. This reward system implicitly motivates the leader to write descriptive sentences about the image, as they will be easily guessed by the other players.

\section{Data and Analysis}
To gather the data, we took ten groups of four volunteers and ran each group on the same ten randomly sampled images from Microsoft's COCO dataset \cite{coco}.
\subsection{People Find SentenceRacer to be more fun}


Qualitative results suggest that SentenceRacer is more fun in comparison to the task of image captioning. Surveys comparing SentenceRacer and a standard AMT image captioning task show that players find SentenceRacer more fun and engaging particularly because of the social and fast-paced aspects of the game.

\subsection{SentenceRacer's Sentences are Confirmed by AMT}
Sentences collected by SentenceRacer were sent to AMT for verification by three crowd workers. A sentence is verified by AMT if at least two out of the three workers agree that the sentence accurately describes the image. A sentence is considered verified by SentenceRacer if all the words in the sentence can be guessed by the players. We found that 87.8\% of sentences verified by SentenceRacer were also verified by AMT, while only 54.9\% of the sentences not verified by SentenceRacer were verified by AMT. We also found that the sentences collected from SentenceRacer have a higher percentage of verified sentences (87.8\%) than those collected from AMT workers (85.5\%) on the same images.

We investigated the relation of the percentage of sentences verified with the number of remaining blanks left in the game. Table 2 shows that the percentage verified increases as the number of blanks decreases. The tail end of the blanks is sparse, causing the data to have high variance. However, we believe that this trend still shows that SentenceRacer's verification process adequately determines whether a sentence accurately describes an image. The number of blank spaces are directly correlated with how likely a sentence will be verified.

\begin{table}[h!]
\centering
\begin{tabular}{|c|c|c|c|c|}
\hline
Source & Total Blanks & \# Sentences & Verified (\%) \\ \hline
& 4 & 7 & 42.80 \\
& 3 & 12 & 50.00 \\
SentenceRacer & 2 & 9 & 33.30 \\
& 1 & 12 & 75.00 \\
& 0 & 49 & \textbf{87.80} \\ \hline \hline
AMT & - & 200 & 85.50 \\ \hline
\end{tabular}
\caption{Correlation between number of blanks and percentage of verified sentences. As the number of blanks decreases, the percentage of sentences verified by AMT increases. Also in comparison, sentences collected from AMT have a lower verification percentage than the sentences collected by SentenceRacer.}
\end{table}

\subsection{SentenceRacer has Higher Sentence Quality than AMT}

Figure 2 shows the quality of some sentences we received from both tasks on AMT and from playing SentenceRacer. We measure sentence quality by the amount of information we can extract from the sentence describing an image. The average number of objects, object attributes, and pairwise-object relationships per sentence is a basic indicator of sentence quality \cite{scenegraph}. 
\begin{figure}[h!]
\centering
\includegraphics[width=3.3in]{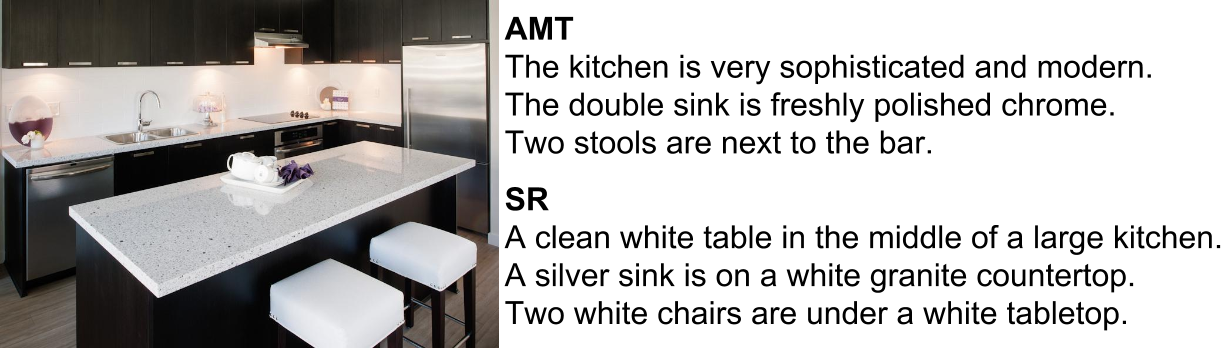}
\caption{Comparing verified sentences from AMT and SentenceRacer .} 
\label{fig-example}
\end{figure}
Table 2 shows that SentenceRacer has statistically significant more objects and relationships and may suggest that SentenceRacer provides more attributes as well. We believe SentenceRacer's sentence quality stems from the rule that correct guesses reward both the guesser and the leader. Players are incentivized to write longer sentences, leading to higher averages of objects, relationships, and attributes, than AMT tasks, where this incentive is absent.

\begin{table}[h!]
\centering
\begin{tabular}{|C{2.5cm}|c|c|c|}
  \hline
& Objects & Relationships & Attributes\\ \hline
\textbf{AMT}\ \ \ \ \ \ \ \ \ \ \ (n=200) & 2.30 & 1.02 & 1.17 \\ \hline
\textbf{SentenceRacer} (n=49) & 2.98 & 1.88 & 1.45 \\ \hline
P-values & $<$ 0.01 & $<$ 0.001 & 0.1 \\ \hline
\end{tabular}
\caption{T-test showing that increased number of objects, relationships, and (potentially) attributes.}
\end{table}

\section{Conclusion}
In this paper, we demonstrate how SentenceRacer is able to collect sentences describing images, an expensive task for Computer Vision research. This system introduces the idea of collecting and verifying sentences through a game that uses contextual cues as a means of entertainment and verification. Our evaluations suggest that this game is more enjoyable than standard methods of collecting sentences. SentenceRacer can also simultaneously perform the collection and verification of sentences. Finally, we show that the sentences collected by SentenceRacer are of higher quality than those collected by AMT. 

We hope to explore how a list of taboo words may increase the diversity of the sentences collected. We also hope to investigate creative ways of using the waiting period between rounds to possibly attempt crowdsourcing other tasks such as grounding objects in the sentence within the image itself.

%
%
%
%
%
\balance

\bibliographystyle{acm-sigchi}
\bibliography{sample}

\begin{thebibliography}{1}

\bibitem{scenegraph}
Johnson, J., Krishna, R., Stark, M., Li, L.-J., Shamma, D.~A., Bernstein, M.,
  and Fei-Fei, L.
\newblock Image retrieval using scene graphs.
\newblock In {\em IEEE Conference on Computer Vision and Pattern Recognition
  (CVPR)} (2015).

\bibitem{coco}
Lin, T.-Y., and et~al.
\newblock {Microsoft COCO: Common objects in context}.
\newblock {\em Computer Vision–ECCV\/} (2014), 740--755.

\bibitem{crossword}
Nickerson, R.~S.
\newblock Five down, absquatulated : Crossword puzzle clues to how the mind
  works.
\newblock {\em Psychonomic Bulletin \& Review\/} (2011), 217--241.

\bibitem{google}
Vinyals, O., Toshev, A., Bengio, S., Erhan, D., and et~al.
\newblock {Show and Tell: A Neural Image Caption Generator}.
\newblock {\em CoRR abs/1411.4555\/} (2014).

\bibitem{gwap1}
von Ahn, L., and Dabbish, L.
\newblock Labeling images with a computer game.
\newblock In {\em Proceedings of SIGCHI}, ACM Press (2001), 9--18.

\bibitem{flickr}
Young, P., Lai, A., Hodosh, M., and Hockenmaier, J.
\newblock {From image descriptions to visual denotations: New similarity
  metrics for semantic inference over event descriptions}.
\newblock {\em Transactions of the Association for Computational
  Linguistics\/}.

\end{thebibliography}

\end{document}